\documentclass[letterpaper,10pt]{article} 

\usepackage{osameet3} 
\usepackage{tikz,pgfplots}
\usepackage{caption}
\usepackage{subcaption}
\usepackage{mathtools}
\usetikzlibrary{calc}
\usepackage{eucal}
\usetikzlibrary{patterns}

\usepackage{amsmath,amssymb}
\usepackage[colorlinks=true,bookmarks=false,citecolor=blue,urlcolor=blue]{hyperref} 

\begin{document}

\title{Experimental Evaluation of Computational Complexity for Different Neural Network Equalizers in Optical Communications}

\author{
  Pedro J. Freire\textsuperscript{(1)}, Yevhenii Osadchuk\textsuperscript{(1)}, Antonio Napoli\textsuperscript{(2)}, Bernhard Spinnler\textsuperscript{(2)} \\ Wolfgang Schairer\textsuperscript{(2)}, Nelson Costa\textsuperscript{(3)}, Jaroslaw E. Prilepsky \textsuperscript{(1)}, Sergei K. Turitsyn\textsuperscript{(1)}
}
\address{  \textsuperscript{(1)}~Aston University, United Kingdom
  \textsuperscript{(2)}~Infinera, Munich, Germany
  \textsuperscript{(3)}~Infinera, Lisbon, Portugal}
\email{p.freiredecarvalhosourza@aston.ac.uk}

\copyrightyear{2021}

 \vspace{-3mm}
\begin{abstract}
  Addressing the neural network-based optical channel equalizers, we quantify the trade-off between their performance and complexity by carrying out the comparative analysis of several neural network architectures, presenting the results for TWC and SSMF set-ups.
\end{abstract}

 \section{Introduction}
The rapid development of neural networks (NN) based signal processing stimulates the research on NN's application for the mitigation of optical transmission impairments ~\cite{Review2020}. In particular, the multi-layer perceptron (MLP), convolutional neural networks (CNNs), and recurrent bidirectional long short-term memory networks (biLSTM) have shown the decent potential to improve the system's performance \cite{freire2021performance}. However, the computational complexity (CC) evaluations for NN-based systems are typically carried out using computational time consumption, which is not a universally indicative parameter. In our work we, first, relate the complexity measures from the machine learning theory to the metrics used in the traditional DSP. Second, we experimentally evaluate the performance enhancement capacity of four different NN architectures (MLP, CNN+MLP, biLSTM, and CNN+biLSTM) and present the detailed study of their CC, addressing both the number of multiplications and the computer run-time. In our study, we use the experimental data from a single channel (SC) 34.4~GBd transmission of a dual-polarization (DP) 16 QAM signal, (i) at 7 dBm launch power through the standard single-mode fiber (SSMF) with $5\times50$~km spans; and (ii) at 2 dBm transmitted through the TrueWave Classic (TWC) fiber over $9\times50$~km spans.

\section{Computational Complexity of Neural Network Equalizers}

For all the NNs, we express the CC in terms of real multiplications per recovered output symbol (RMpS) ~\cite{spinnler2010equalizer}. The offline training CC (the calibration stage) is not taken into account because we will be evaluating the real-time CC (referring to the evaluation stage), which is the most critical parameter for the hardware device's realization. Early works presented some particular results regarding the complexity of MLP~\cite{sidelnikov2018equalization}, RNN~\cite{zhou2019low}, and LSTM~\cite{9324921} layers.  In this work, we directly relate those complexities to the parameters of the most widely used machine learning platforms (Keras, TensorFlow, and PyTorch) without losing generality, and specifically address the most efficient composite NN equalizer types. 

\vspace{-4mm}
\begin{table*}[ht!]
 \centering
\caption{CC per layer for LSTM, 1D-CNN and Dense (MLP) architectures. For more details see \cite{freire2021performance}.} \label{tab:table1}
\vspace{-2mm}
\resizebox{0.75\textwidth}{!}{%
\begin{tabular}{|c|c|c|c|c|}
\hline
NN Layer & Input Shape & Output Shape & Number of RMpI& Big-O \\ \hline
LSTM & $[B,n_{s},n_{i}]$ & $[B,n_{s},n_{o}]$ & $ n_{s}n_{h}(4n_i+4n_{h}+3+n_o)$ & $\mathcal{O}(n d^2)$ \\ \hline
1D-CNN & $[B,n_{s},n_{i}]$ & $[B,n_{s'},n_{o}]$ & $ k n_{i} n_{o} n_{s'}$ & $\mathcal{O}(knd^2)$ \\ \hline
Dense (MLP) & $[B,n_{i}]$ & $[B,n_{o}]$ & $ n_{i}n_{1}+n_{1}n_{o}$ & $\mathcal{O}(nd)$ \\ \hline
\end{tabular}}
\end{table*}%

Let $B$ be the mini-batch size; $n_s$ -- the size of the input time sequence, with $n_s=M$, where $M$ is the memory size of the model; $n_{i}$ -- the number of features; and $n_o$ -- the number of outputs per symbol. The number of real multiplications per input (RMpI) for CNN, LSTM, and Dense (MLP) layers is summarized in Table~\ref{tab:table1}: we show the number of multiplications as a function of the input and output shapes of the layers and their design. The design refers to: the number of filters $f$ and kernel size $k$ for CNN; the number of hidden units $n_h$ for LSTM; and the number of hidden neurons $n_1$ for one MLP layer. In the case of the CNN, the variable $n_{s'}$ is defined as:
\vspace{-1mm}
\begin{equation}\nonumber
n_{s'} = \frac{n_s + 2 \, padding -dilation \, (k-1)-1}{stride} +1{.}
\label{eq:outconv}
\end{equation}\vspace{-1mm}
We also show the complexity of the studied NNs in the last column of Table~\ref{tab:table1} in Big-O notation, where $n$ is the sequence length, and $d$ is the representation dimension. Table \ref{tab:table1} presents the correct amount of operations for a single layer, but it lacks the details for multi-layered structures that can cause the CC misestimation when building composite equalizers, e.g., due to counting twice the number of multiplications between the input and output of two sequential layers \cite{freire2021performance}. In this paper we analyze the results for:  i) 3 hidden dense layers (MLP), ii) 1d-CNN + 2 hidden layers, iii) 1 biLSTM layer, and iv) 1d-CNN + 1 biLSTM layer. We use flattening layers when reducing the dimensionality of the data. The codes of the NN equalizers used in this paper are available in Ref.~\cite{freire2021performance}.

Let us explain the input/output shapes of the equalizers. For biLSTM and CNN layers, the input is parametrized as $[B, n_s, n_i]$, the three numbers defining the dimensions of the input tensor. The parametrization for the MLP equalizer is simpler, with $[B,n_s \cdot n_i]$ defining the dimensions of the 2D tensor input. For all equalizers, the output shape is $[B,n_o]$. Since we use the real and imaginary parts for both polarizations as input to recover, the real and imaginary parts of a symbol in each of the polarization, $n_{i}$ and $n_o$ are $4$ and $2$, respectively.  With this in mind, we have used the EQs. (8), (9), (12), and (13) from Ref.~\cite{freire2021performance} to calculate the CC for each NN equalizer.

We emphasize that the true CC metric is the processing time, or the latency of the equalizer, that depends on both software implementation and hardware selection. Nevertheless, the RMpS metric provides a good estimate of the CC that depends on nothing but the chosen NN topology. To demonstrate this, we show in Fig.~\ref{fig:Result_02}~(a) the averaged latency (inference time) for one equalized symbol at four different complexity levels, ranging from $10^4$ to $10^7$ RMpS, for four distinct NN designs, when using Colab CPU (4vCPU at 2.0GHz with 26GB RAM). For the same degree of RMpS, the MLP and CNN+MLP architectures display similar results. However, when using LSTM layers, the increase of processing time of up to $0.6$~ms can be observed in the case of using highly complex structures. This effect is the consequence of Keras' implementation of the LSTM layer, in which the input data is subjected to additional pre-processing operations to construct the states of each recurrent cell. The difference is not noticeable when using CNN or MLP only, because the inference is essentially a simple feed-forward matrix multiplication operation.  Fig.~\ref{fig:Result_02}~(a) confirms that the computation of RMpS is a good method for the CC evaluation since we obtained nearly identical results in terms of processing time using different NN designs with similar RMpS. For instance, the latency per recovered symbol for the NN equalizers of $10^4$ RMpS was around $0.02$~ms, and when we used the topologies with $10^6$ RMpS, the latency increased to up $0.25$~ms per symbol.

Finally, we note that the number of trainable parameters determines the memory required by the equalizer, but not the CC of the NN model. For example, the CNN+MLP, MLP, biLSTM, and CNN+biLSTM equalizers with $10^7$ RMpS have similar processing time, but the first one has $10.5$M trainable parameters, the second -- $11.6$M, the third -- $370$k, and the last one has $125$k.


\section{Experiment setups}
At the transmitter side, the data bits were generated using a PRBS of order 32, then mapped to 16-QAM and filtered digitally using an RRC filter with the roll-off 0.1 to limit the channel bandwidth to 37.5~GHz. The samples were uploaded to a digital-to-analog converter, the output of which was amplified by an electrical amplifier that drives a DP I/Q Mach-Zehnder modulator. We studied two configurations: $5\times 50$~km spans of SSMF and $9\times 50$~km spans of TWC fiber.  The parameters of the TWC fiber are: the attenuation parameter $\alpha = 0.23$ dB/km, the dispersion coefficient $D = 2.8$ ps/(nm$\cdot$km), and the effective nonlinearity coefficient $\gamma = 2.5$~(W$\cdot$km)$^{-1}$. The SSMF parameters are: $\alpha = 0.2$ dB/km, $D = 17$ ps/(nm$\cdot$km), and $\gamma = 1.2$ (W$\cdot$km)$^{-1}$. Each span was followed by an EDFA with a noise figure $\sim 4.5$~dB. At the RX, the signal was detected using an integrated coherent receiver. Then, the balanced photodiodes performed the optical-electrical conversion. The electrical signal was sampled by an ADC and processed using a traditional DSP ~\cite{freire2020complex}. The symbols at the DSP output were processed by an NN-based equalizer. The training and testing dataset were generated independently with a cross-correlation $ < 0.02$. 

\vspace{-3mm}
\section{Results and discussions}

We consider the CCs in the range $10^{3}$ to $10^{8}$. The hyper-parameters distributions for each NN architecture are given in Ref.~\cite{pedro_jorge_freire_2021_4836177}. The parameters of those topologies were also tuned by the Bayesian optimizer (BO) \cite{freire2020complex}. However, the BO search range was reduced to comply with each CC constraint.
\begin{figure}[ht!] 
 \begin{subfigure}[t]{.33\textwidth}
  \centering
\begin{tikzpicture}[scale=0.67]
  \begin{axis}[
    major x tick style = transparent,
    ybar=2*\pgflinewidth,
    bar width=7pt,
    ymajorgrids = true,
    y label style={at={(axis description cs:0.11,.5)},rotate=0,anchor=south},
    ylabel={Processing Time [ms]}, 
    xlabel={\#Real Multiplications per symbol},
    symbolic x coords={ $10^4$, $10^5$, $10^6$, $10^7$},
    xtick = data,
    scaled y ticks = true,
    enlarge x limits=0.25,
    ymin=0,
    ymax=1.0,
        legend pos=north west,
                ]
    \addplot[style={black,fill=red,mark=none},
  postaction={
    pattern=horizontal lines}]
      coordinates {($10^4$,0.00923 ) ($10^5$,0.0207) ($10^6$, 0.0757) ($10^7$,0.37)};
    \addlegendentry{\footnotesize{MLP}};

    \addplot[style={black,fill=green,mark=none},
  postaction={
    pattern=vertical lines}]
      coordinates {($10^4$,0.0123) ($10^5$,0.0291) ($10^6$,0.09508) ($10^7$,0.37)};
    \addlegendentry{\footnotesize{CNN+MLP}};
    
    \addplot[style={black,fill=orange,mark=none},
  postaction={
    pattern=dots}]
      coordinates {($10^4$,0.0377) ($10^5$,0.0657) ($10^6$,0.19) ($10^7$,0.728)};
    \addlegendentry{\footnotesize{CNN+biLSTM}};
    
    \addplot[style={black,fill=blue,mark=none},
  postaction={
    pattern=north east lines}]
      coordinates {($10^4$,0.0330) ($10^5$,0.0612) ($10^6$,0.25) ($10^7$,0.971) };
    \addlegendentry{\footnotesize{biLSTM}};
    
    \legend{MLP,CNN+MLP,CNN+biLSTM,biLSTM}
  \end{axis}
\end{tikzpicture}
   \caption{}
  \label{experiment_result1} 
 \end{subfigure}
  \begin{subfigure}[t]{.33\textwidth}
  \centering
 \begin{tikzpicture}[scale=0.67]
  \begin{axis} [
    grid=both,
    xmode=log, 
    xmin=1e3,
    xmax=1e8,
  	  ymajorgrids = true,
  grid style = dashed,
    y label style={at={(axis description cs:0.12,.5)},rotate=0,anchor=south},
    ylabel={Q-factor Gain [dB]}, 
    xlabel={\#Real Multiplications per symbol},
    legend style={legend pos=north west, legend cell align=left,fill=white, fill opacity=0.6, draw opacity=1,text opacity=1},
  	grid style={dashed}]
    ]

  \addplot[color=red, mark=o,, very thick]  
  coordinates {
  (1000,0.37) (10000,0.62) (100000,0.98) (1000000,0.93) (10000000,0.82) (100000000,0.82)

  };
  \addlegendentry{\footnotesize{MLP}};

  \addplot[color=blue, mark=square, very thick]  
  coordinates {
  (1000,0.32) (10000,0.35) (100000,0.68) (1000000,1.02) (10000000,1.85)  (100000000,1.95)

  };
  \addlegendentry{\footnotesize{biLSTM}};

      \addplot[color=orange, mark=star, very thick]  
  coordinates {
  (1000,-0.28) (10000,0.36) (100000,0.54) (1000000,0.92) (10000000,1.72) (100000000,1.99)

  };
  \addlegendentry{\footnotesize{CNN+biLSTM}};

      \addplot[color=green,mark=diamond, very thick]  
  coordinates {
  (1000,-0.61) (10000,0.56) (100000,0.89) (1000000,1.25) (10000000,1.48) (100000000,1.58) 

  };
  \addlegendentry{\footnotesize{CNN+MLP}};
  \end{axis}
    \node[text width=2cm] at (1.5,1.9) 
  {T.1};
      \node[text width=2cm] at (2.5,3.1) 
  {T.2};
      \node[text width=2cm] at (3.8,2.05071082) 
  {T.3};
      \node[text width=2cm] at (5.0,4.35) 
  {T.4};
      \node[text width=2cm] at (6.4,5.45) 
  {T.5};
        \node[text width=2cm] at (7.4,5.45) 
  {T.6};
              \node[text width=2cm] at (2.05,0.2) 
    {DBP Region};
\draw[pattern=north west lines, pattern color=blue] (0,0) rectangle (0.5, 3.5);
  
  \end{tikzpicture}
   \caption{}
  \label{experiment_result1} 
 \end{subfigure}
 \begin{subfigure}[t]{.33\textwidth}
  \centering
 \begin{tikzpicture}[scale=0.67]
  \begin{axis} [
    y label style={at={(axis description cs:0.12,.5)},rotate=0,anchor=south},
    ylabel={Q-factor Gain [dB]}, 
    xlabel={\#Real Multiplications per symbol},
    grid=both,
    xmode=log, 
    xmin=1e3,
    xmax=1e8,
  	  ymajorgrids = true,
  grid style = dashed,
    legend style={legend pos=north west, legend cell align=left,fill=white, fill opacity=0.6, draw opacity=1,text opacity=1},
  	grid style={dashed}]
    ]
  
  \addplot[color=red, mark=o,, very thick]  
  coordinates {
  (1000,0.42) (10000,0.93) (100000,1.92) (1000000,1.75) (10000000,1.71) (100000000,1.69)

  };
  \addlegendentry{\footnotesize{MLP}};

  \addplot[color=blue, mark=square, very thick]  
  coordinates {
  (1000,0.13) (10000,0.27) (100000,0.78) (1000000,1.36) (10000000,2.70)  (100000000,2.76)

  };
  \addlegendentry{\footnotesize{biLSTM}};
  

  
      \addplot[color=orange, mark=star, very thick]  
  coordinates {
  (1000,0.16) (10000,0.49) (100000,1.14) (1000000,1.85) (10000000,2.85) (100000000,2.80)

  };
  \addlegendentry{\footnotesize{CNN+biLSTM}};

      \addplot[color=green,mark=diamond, very thick]  
  coordinates {
  (1000,0.3) (10000,0.48) (100000,1.06) (1000000,1.82) (10000000,2.2) (100000000,2.07) 

  };
  \addlegendentry{\footnotesize{CNN+MLP}};
  \end{axis}
    \node[text width=2cm] at (1.5,1.5) 
  {T.1};
      \node[text width=2cm] at (2.3,2.45) 
  {T.2};
      \node[text width=2cm] at (3.9,3.071082) 
  {T.3};
      \node[text width=2cm] at (5.0,4.1) 
  {T.4};

      \node[text width=2cm] at (6.4,5.45) 
  {T.5};
        \node[text width=2cm] at (7.4,5.45) 
  {T.6};
            \node[text width=2cm] at (2.05,0.2) 
    {DBP Region};
\draw[pattern=north west lines, pattern color=blue] (0,0) rectangle (0.5, 2.5);
  \end{tikzpicture}
   \caption{}
  \label{experiment_result2} 
 \end{subfigure} 
  \vspace{-5mm}
 \caption{ (a): Processing time to equalize one symbol vs. RMpS comparison.  (b) and (c): Q-factor gain dependence on RMpS for different NN architectures, (b) -- for the SSMF 5$\times$50~km at 7~dBm, and (c) -- for the TWC fiber with 9$\times$50~km at 2~dBm.}\label{fig:Result_02}
\end{figure}
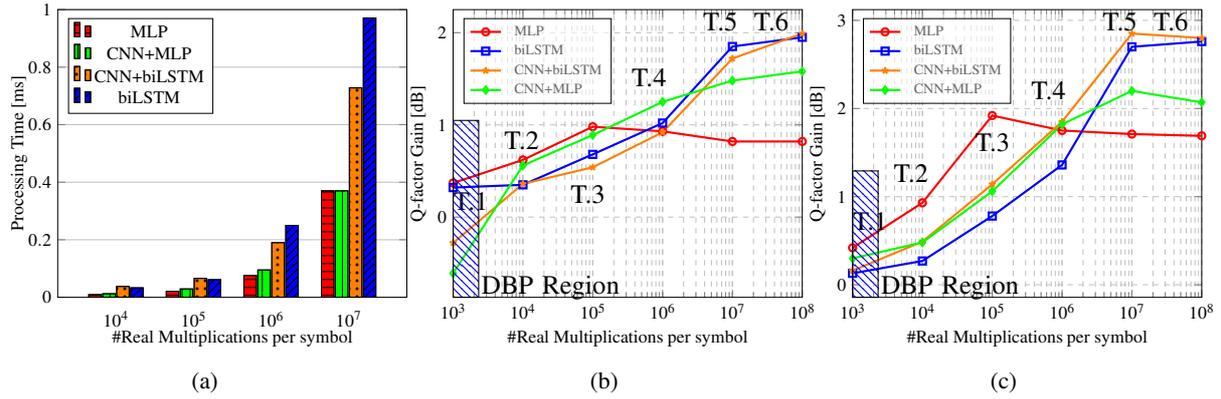
Fig.~\ref{fig:Result_02} (b) and (c) depict the performance gain (measured against the non-equalized case) of the different equalizers for different CC levels. We can see that the CC impacts each equalizer differently. For both the SSMF and TWC fiber, the CNN+biLSTM, CNN+MLP, and biLSTM equalizers need $ \geq 10^7$ RMpS to achieve the best performance. The highest Q-factor gain is achieved by CNN+biLSTM, which is only marginally better than the biLSTM equalizer.
The MLP only shows the worst performance. Fig.~\ref{fig:Result_02} also shows that we have similar trends for both fiber types studied. In the TWC scenario, the CNN+biLSTM architecture provided the highest Q-factor improvement for high complexity ($ \geq 10^6$ RMpS), whereas the MLP provided the best performance at lower complexity ($ \approx10^5$ RMpS). This tendency can be explained by the fact that learning and unrolling the nonlinear dynamics of the data symbols requires more filters and a bigger number of hidden units in the combined architectures. Since the complexity of recurrent and convolutional layers is significantly higher than that for dense layers, Table \ref{tab:table1}, the CC restriction severely limits the design choices for such layers. Similar conclusions can be drawn for SSMF. In this case, the biLSTM equalizer outperforms the CNN+biLSTM one except for the highest complexity level. This result is a consequence of the higher chromatic dispersion of SSMF which requires more taps for its compensation. Thus, for the same overall CC, fewer filters are used in the CNN+biLSTM equalizer, which reduces its performance. For the lower CC, the MLP still shows the best performance. However, the CNN+MLP also shows quite good performance when using the TWC fiber.  Finally, Fig.~\ref{fig:Result_02} shows that increasing the CC does not necessarily lead to performance improvement (e.g., when using the MLP). This effect is known in machine learning as the model capacity limit \cite{Goodfellow-et-al-2016}, where the model begins to overfit after a certain amount of capacity is reached by the model design.

The hatched blue zone in the panels demonstrates the results for the  DBP with 3 StPS, to contrast the performance/complexity of the NN equalizers to the DBP. Then, it is evident that reducing the number of neurons, filters, and hidden units, does not give us low complexity architectures, because the performance falls below the DBP level. In both scenarios, Topologies 3, 4, and 5 revealed better results compared to the DBP, but at the expense of much higher CC.  As a possible alternative,  pruning and quantization techniques \cite{cheng2017survey} can be used to minimize the computational complexity of the NN equalizers without compromising their performance.


\section{Conclusions}
We compared the performance and complexity of NN-based equalizers using experimental data for two benchmark systems: the SSMF with $5 \times 50$ km spans and the TWC with $9 \times 50$ km spans, each with 34.4 GBd 16-QAM DP. We showed that the best-performing architecture type crucially depends on the complexity level and transmission scenario. Additionally, we advocate the CC as an indicator of the real multiplicative complexity for NN equalizers.


\section{Acknowledgements}

\footnotesize
\linespread{0.0}
This work has received funding from: EU Horizon 2020 program under the MSCA grant agreement No. 813144 (REAL-NET) and the support of Leverhulme Trust and  TRANSNET.
\normalsize
\linespread{1.0}

\end{document}